\documentclass[aps,pra,onecolumn,showpacs,superscriptaddress]{revtex4}
\usepackage{bm,amsmath,amssymb,latexsym,mathrsfs,graphicx,enumerate}
\usepackage[mathcal]{euscript}
\usepackage{hyperref}
\usepackage{epsfig}

\begin{document}

\title{\textbf{Solitary electromagnetic waves propagation in the asymmetric oppositely-directed coupler}}
\author{Elena V. Kazantseva }
\email{kazantsevaev@ornl.gov} \affiliation{Center for Engineering
Science Advanced Research, Computer Science and Mathematics
Division, Oak Ridge National Laboratory, Oak Ridge, TN 37831, USA}
\author{Andrei I. Maimistov }
\email{maimistov@pico.mephi.ru}\affiliation{Department of Solid
State Physics and Nanosystems, Moscow Engineering Physics Institute,
Moscow 115409, Russia}
\author{Sergei S. Ozhenko}
\email{ozhenko@gmail.com}\affiliation{Department of Solid State
Physics and Nanosystems, Moscow Engineering Physics Institute,
Moscow 115409, Russia}

\begin{abstract}
\noindent We consider the electromagnetic waves propagating in the
system of coupled waveguides. One of the system components is a
standard waveguide fabricated from nonlinear medium having positive
refraction and another component is a waveguide produced from an
artificial material having negative refraction. The metamaterial
constituting the second waveguide has linear characteristics and a
wave propagating in the waveguide of this type propagates in the
direction opposite to direction of energy flux. It is found that the
coupled nonlinear solitary waves propagating both in the same
direction are exist in this oppositely-directed coupler due to
linear coupling between nonlinear positive refractive waveguide and
linear negative refractive waveguide. The corresponding analytical
solution is found and it is used for numerical simulation to
illustrate that the results of the solitary wave collisions are
sensible to the relative velocity of the colliding solitary waves.
\end{abstract}








\maketitle

\section{Introduction}

\noindent The waveguide structure fabricated from the two closely
placed waveguides is of common use in the fiber and integrated
optics. Coupling between the waveguides is due to tunnel penetration
of light from one waveguide into another waveguide
\cite{YarivYeh:84,Yariv:73}. This coupler preserves direction of
light propagation, and for this reason it is named a directed
coupler. It was found \cite{FSOASS:87,HWS:88} that the steady state
pair of electromagnetic pulses can exist in the extended directed
coupler or twin-core fibers \cite{Hasegawa}. Sometimes it termes
soliton. There is a large body of publications devoted to
investigation of the soliton generation and its propagation, see for
example \cite{4,5,6,7}.

Recent progress in nanofabrication has lead to the design of new
materials with highly unusual optical properties
\cite{bib1,bib2,bib3,bib4,bib5,10,11,12}, of which\textit{negative
refraction} is an example. Negative refraction occurs in media in
which the wave vector of the electromagnetic wave is antiparallel to
the Poynting vector \cite{bib6,bib7,bib8}. In the particular case
when the real parts of the dielectric permittivity and magnetic
permeability in the medium simultaneously take on negative values in
some frequency range, the property of negative refraction will
appear. The existence of such media was demonstrated experimentally
first in the microwave and then in the near-infrared ranges
\cite{bib2,bib3,bib4,bib5,b20,b21}. Negative refraction can be
realized for the media with spatial dispersion \cite{12} which is
large enough and in the photonic crystals \cite{16,17,18}. Recently
bulk metamaterials that show negative refraction for all incident
angles in the visible region were presented \cite{Zhang1,Zhang2}.

The unusual properties of negative refractive (NR) index materials
reveal themselves most prominently when the refractive index of the
same medium can be positive in one spectral region and negative in
another \cite{bib8,Kivsh,Popov:06,Popov:06a}. New wave propagation
phenomena can also be expected when a wave passes through, or is
localized near, an interface between such a material and a
conventional dielectric, i.e., positive refractive (PR) index
material \cite{Shadr:04,Darmanyan:05,Darmanyan:07,ZSZK:05}. We can
refer to these cases as \textit{negative-positive refraction} ones.
The intriguing example of the negative-positive refraction medium is
the coupler, where one of the waveguides is fabricated from a
material with a negative refractive index. This device acts as a
(distributed) mirror. The radiation entering one waveguide leaves
the device through the other waveguide at the same end but in the
opposite direction. For this reason, this device can be called a
\textit{oppositely-directed coupler} (ODC). The ODC is known to
support the propagation of linear waves with a gap in their spectrum
(forbidden zone). It has been shown recently~\cite{LGM:07} that due
to this gap the \textit{nonlinear} oppositely directed coupler is
bistable. Bistability results from the multi-valued dependence of
the transmission coefficient  on the input-wave power. It is
noteworthy that this effect has no analogy in conventional directed
couplers consisting of uniform waveguides without a mirror-based
feedback mechanism. It was found \cite{MGL:09} that if the waveguide
nonlinear optical properties are characterized by the third order
susceptibility, a coupled pair of the steady state pulses (each one
is localized in its own waveguide) can exist in ODC.

The fabrication of the transparent nonlinear NR waveguide is a
complicated and still unresolved problem. The linear NR waveguide
could be more suitable for realization. However there is question is
it possible a quasisolitonic regime for the waves in such a
\textit{antisymmetric} coupler consisting of a nonlinear
conventional dielectric waveguide coupled with a linear NR
waveguide. In this paper we consider the extended asymmetric
oppositely-directed coupler (AODC). We found the steady state
solutions of the equations which describe the waves in the AODC. The
numerical simulation of these steady state pulses formation and the
collisions between the quasisolitons is performed. We found that the
pulses are very robust against perturbations. It allows a definite
conclusion that the steady state pulses are a new kind of gap
solitons. It should be noted that this solitary wave exists in the
medium without grating. The influence of linear losses in the NR
waveguide on the quasisoliton formation discussed shortly.

\section{The model formulation}

\noindent We consider the waveguide structure in which one of the
channels is made from nonlinear material features by the positive
refraction, and the NR channel is composed of linear material. The
system of the equations based on connected wave approximation takes
the form
\begin{eqnarray}
i  \frac{\partial q_{1}}{\partial z}&+&
\frac{i}{v_{g1}}\frac{\partial q_{1}}{\partial t} + K_{12}q_2
e^{i\Delta \beta z} + \nonumber \\
&+& \frac{2\pi \omega_0}{c} \sqrt{\frac{\mu_1(\omega_0)}{\epsilon_1(\omega_0)}}A^2_{0}\chi^{(3)}_{eff}|q_1|^2q_1  = 0, \label{eq1m}   \\
-i \frac{\partial q_{2}}{\partial z} &+& \frac{i}{v_{g2}}\frac{\partial
q_{2}}{\partial t} + K_{21}q_1 e^{-i \Delta \beta z} = 0.
\nonumber
\end{eqnarray}
The coefficient $\chi^{(3)}_{eff}$ is the effective non-linear
susceptibility of the first channel. $K_{12}$ and $K_{21}$ are the
coupling constants. Linear properties of the PR channel defined by
dielectric permittivity $\varepsilon_{1}$ and magnetic permeability
$\mu_{1}$. These values are assumed be real, that corresponds to the
lossless materials. Slowly varying envelope of the electric field
$E_{j}(z,t)$ in $j$-th channel is written as $E_{j}(z,t)= A_0
q_j(z,t)$, $v_{gj}$ is the group-velocity of the wave in $j$-th
channel, $(j=1,2)$. Here we assume that the group-velocity
dispersion can be neglected. In the following we also assume that
the synchronism condition is satisfied: $\Delta \beta=0$. It is
convenient to use new normalized variable
\begin{eqnarray*}
Q_1 = \sqrt{K_{21}}q_1 e^{-i\Delta \beta z}, ~~ Q_2 =
\sqrt{K_{12}}q_2 e^{i\Delta \beta z}, \\
\zeta = z/L_c, ~~~ \tau = t_0^{-1}(t-z/V_0), ~~~ L_c =
(K_{12}K_{21})^{-1/2}, \\
t_0=L_c(v_{g1}+v_{g2})/2 v_{g1}v_{g2}, ~~~  V_0^{-1}=
(v_{g1}-v_{g2})/2 v_{g1}v_{g2}.
\end{eqnarray*}
In the normalized variables the ultimate system of equations
(\ref{eq1m}) for the AODC reads as

\begin{equation}
\label{eq2}
\begin{array}{rcl}
i \left(  \frac{\partial}{\partial
\zeta}+\frac{\partial}{\partial\tau}\right)Q_1+Q_2+r|Q_1|^2Q_1&=&0, \\
i \left(  \frac{\partial}{\partial
\zeta}-\frac{\partial}{\partial\tau}\right)Q_2-Q_1&=& 0.
\end{array}
\end{equation}
The parameter of nonlinearity is
\[ r=\frac{2\pi \omega_0}{c
K_{21}\sqrt{K_{12}K_{21}}}\sqrt{\frac{\mu_1(\omega_0)}{\epsilon_1(\omega_0)}}A^2_{0}\chi^{(3)}_{eff}.
\]

Let us consider the linear wave limit of the AODS equations. From
(\ref{eq2}) one can get
\begin{equation}
\label{eq2lin}
\begin{array}{rcl}
i \left(  \frac{\partial}{\partial
\zeta}+\frac{\partial}{\partial\tau}\right)Q_1+Q_2 &=&0, \\
i \left(  \frac{\partial}{\partial
\zeta}-\frac{\partial}{\partial\tau}\right)Q_2-Q_1&=& 0.
\end{array}
\end{equation}
If we take the Fourier transformation
\[
Q_{1,2}=\int^{+\infty}_{-\infty}\tilde{Q}_{1,2}e^{-i\nu \tau
+i\kappa \zeta }\frac{d\kappa d\nu}{4\pi^2},
\]
the equations (\ref{eq2lin}) result in the following linear system
of equation
\begin{eqnarray*}
(\nu - \kappa)\tilde{Q}_1 + \tilde{Q}_2 =0, \\
(\nu + \kappa)\tilde{Q}_2 + \tilde{Q}_1 =0.
\end{eqnarray*}
This system of equations has the nonzero solution only if the
corresponding determinant
\[
\det \left (
\begin{array}{cc}
\nu - \kappa & 1 \\
1& \nu + \kappa
\end{array}
\right)
\]
is equal to zero. That leads to the dispersion relation
\begin{equation}
\nu (\kappa) = \pm \sqrt{1+\kappa ^2}. \label{disprel}
\end{equation}
Thus the spectrum of the linear waves has the gap $\Delta \nu _g=
2$. This gap is characteristic feature for a distributed mirror
\cite{Yeh}. Hence, the AODC in linear wave limit acts as a mirror.

In following consideration it is suitable to take the real variables
form of equations (\ref{eq2}). By using the real variables
$Q_{1,2}=a_{1,2}e^{i\phi_{1,2}}$ one obtains
\[
\left(  \frac{\partial}{\partial
\zeta}+\frac{\partial}{\partial\tau}\right)a_1=a_2 \sin
\Phi,~~~\left( \frac{\partial}{\partial
\zeta}-\frac{\partial}{\partial\tau}\right)a_2=a_1 \sin \Phi,
\]
\begin{equation}
\label{eq3} \\
\left(  \frac{\partial}{\partial
\zeta}+\frac{\partial}{\partial\tau}\right)\phi_1=\frac{a_2}{a_1}\cos\Phi+ra^2_1,
\end{equation}
\[
\left(  \frac{\partial}{\partial
\zeta}-\frac{\partial}{\partial\tau}\right)\phi_2=-\frac{a_1}{a_2}\cos\Phi,
\]
where $\Phi=\phi_1-\phi_2$. From the amplitude equation it follows
that
\[
\frac{\partial}{\partial\zeta}\left(a^2_2-a^2_1\right)=\frac{\partial}{\partial\tau}\left(
a^2_1+a^2_1 \right).
\]
Hence,
\begin{equation}
\frac{\partial}{\partial\zeta}\int\limits^{+\infty}_{-\infty}\left(a^2_2-a^2_1\right)d\tau=
\left. \left( a^2_1+a^2_1 \right)
\right|^{+\infty}_{-\infty}.\label{eq4}
\end{equation}
In the case of solitary waves,for which the electromagnetic fields
vanish at infinity are considered, the right part of (\ref{eq4}) is
equal to zero. It leads to integral of motion
\begin{equation}
\int\limits^{+\infty}_{-\infty}\left(a^2_2-a^2_1\right)d\tau =
\textrm{const}.\label{intmot}
\end{equation}
It is the modified Manley-Rowe relation. As usually, in the case of
the quadratic nonlinear PR media Manley-Rowe relation looks like
\[
\int\limits^{+\infty}_{-\infty}\left(a^2_2+a^2_1\right)d\tau =
\textrm{const}.
\]
The difference of the expression (\ref{intmot}) from conventional
Manley-Rowe relation is explained by the fact that the energy flux
for the waves in one waveguide opposite in direction to energy flux
of other waveguide, while their wave vectors are approximately the
same. This flux pattern is an inherent feature of the
negative-positive refraction media.

\section {Analytical solution --- the steady-state pulse}

\noindent To consider the solitary steady state waves in AODC by
similar way as in \cite{MGL:09} we have start from the equations
(\ref{eq3}). Suppose that solutions of these equations are depend
only on single variable
$$\eta=\frac{\zeta+\beta\tau}{\sqrt{1-\beta^2}}$$ with free
parameter $\beta$. Suppose $u_1=\sqrt{1+\beta a_1}$ and
$u_2=\sqrt{1-\beta a_2}$. The system of the equations (\ref{eq3})
takes the following form
\begin{equation}
\frac{\partial}{\partial \eta}u_1 = u_2 \sin \Phi,~~~\frac{\partial}{\partial \eta}u_2 = u_1 \sin \Phi,  \label{eq5a} \\
\end{equation}
\begin{eqnarray}
\frac{\partial}{\partial \eta}\phi_1&=&\frac{u_2}{u_1}\cos\Phi+\Theta u^2_1, \label{eq5b}\\
\frac{\partial}{\partial \eta}\phi_2&=&-\frac{u_1}{u_2}\cos\Phi,
\label{eq5c}
\end{eqnarray}
where $$\Theta=\frac{r}{1+\beta}\sqrt{\frac{1-\beta}{1+\beta}}.$$ We
can also write an equation for the phase difference
\begin{equation}
\frac{\partial}{\partial\eta}\Phi=\left(\frac{u_1}{u_2}+\frac{u_2}{u_1}\right)\cos\Phi+\Theta
u^2_1.\label{eq6}
\end{equation}
The phase equations can be used to
get one equation for the phase difference. Finally, the total system
of equations reads
\begin{eqnarray}
\frac{\partial}{\partial \eta}u_1&=&u_2 \sin \Phi,\label{eq7a}\\
\frac{\partial}{\partial \eta}u_2&=&u_1 \sin \Phi,  \label{eq7b} \\
\frac{\partial}{\partial
\eta}\Phi&=&\left(\frac{u_1}{u_2}+\frac{u_2}{u_1}\right)\cos\Phi+\Theta
u^2_1.\label{eq7c}
\end{eqnarray}

We are looking for a solution in a form of the solitary wave, it
corresponds with the following boundary condition
$$a_{1,2}\rightarrow 0  ~~\textrm{at} ~~  \eta \rightarrow \pm \infty.$$
From the equation (\ref{eq7a}) and (\ref{eq7b}) it follows
$u^2_1=u^2_1$, or $u_1=\epsilon u_2$, where $\epsilon=\pm 1$. Hence,
the system of equations (\ref{eq7a})-(\ref{eq7c}) is reduced to
following pare of equations
\begin{eqnarray}
\frac{\partial}{\partial \eta}u_1&=&\epsilon u_1 \sin \Phi,\label{eq8a}\\
\frac{\partial}{\partial \eta}\Phi&=&2\epsilon\cos\Phi+\Theta u^2_1.
\label{eq8b}
\end{eqnarray}
Multiplying the last equation by $a^2_1\sin\Phi$ and taking into
account the equation (\ref{eq8a}), we  get the second integral of
motion
\begin{equation}
u^2_1\left(\cos\Phi+\frac{\epsilon\Theta}{4}u^2_1\right)=C_2.
\label{eq9}
\end{equation}
Due to the boundary condition, the value of this integral is equal
to zero. As we are looking for a non zero solution of (\ref{eq8a})
and (\ref{eq8b}), one can write
\begin{equation}
\cos\Phi+\frac{\epsilon\Theta}{4}u^2_1=0.
\label{eq10}
\end{equation}
Substitution of (\ref{eq10}) into (\ref{eq8a}) leads to
$$(du_1
/d\eta )^2=u^2_1\left(1-(\Theta/4)^2u^4_1\right).
$$
Choosing the variable $u_1=w^{-1/2}$ reduces this expression into
equation
$$(dw /d\eta )^2=4\left(w^2-(\Theta/4)^2\right),
$$
which has the following solution
\begin{equation}
w(\eta)=(\Theta/4)\cosh2(\eta-\eta_2).\label{eq11}
\end{equation}
Thus, the solution of (\ref{eq8a}) and (\ref{eq8b}) is
$$
u^2_1(\eta)=u^2_2(\eta)=\frac{4/\Theta}{\cosh2(\eta-\eta_2)}.
$$
By using the phase's equations and (\ref{eq11}) one can write
$$
\frac{\partial}{\partial\eta}\phi_1=\frac{3}{4}\Theta
u^2_1(\eta)=\frac{3}{\cosh2(\eta-\eta_2)} ,
$$
$$
\frac{\partial}{\partial\eta}\phi_2=\frac{1}{4}\Theta
u^2_1(\eta)=\frac{1}{\cosh2(\eta-\eta_2)}
$$
that yields
\begin{eqnarray}
\phi_1(\eta)&=&\phi_1(-\infty)+3\arctan\left(\exp2(\eta-\eta_2)\right),
\label{eq12}\\
\phi_2(\eta)&=&\phi_2(-\infty)+\arctan\left(\exp2(\eta-\eta_2)\right).
\end{eqnarray}

The phases $\phi_{1,2}(-\infty)$ should be chosen in such a way that
(\ref{eq10}) is satisfied. As the solitary wave's amplitude tends to
zero $u_{1,2}(\infty)\rightarrow 0$, the relation (\ref{eq10}) at
infinity reduces to
$$\cos\Phi(-\infty)=\cos(\phi_1(-\infty)-\phi_2(-\infty))=0.$$
In the numerical simulation we choose $\phi_1(-\infty)=0$ and
$\phi_2(-\infty)=-\pi/2$.

The amplitudes $a_{1,2}$ are defined by the following expressions
\begin{eqnarray}
a^2_1(\eta)=\frac{4}{\Theta(1+\beta)\cosh2(\eta-\eta_2)},\label{eq13a}\\
a^2_2(\eta)=\frac{4}{\Theta(1-\beta)\cosh2(\eta-\eta_2)}.
\label{eq13b}
\end{eqnarray}

The expressions (\ref{eq13a}) and (\ref{eq13b}) describe the steady
state solitary wave propagating in AODC under consideration. This
wave looks like a gap soliton propagating in the nonlinear Bragg
grating, however there are no any periodic structures. Gap in the
linear wave spectrum (\ref{disprel}) is due to the flux pattern in
this negative-positive refraction media.

\begin{figure}
\centerline{ \epsfig{file=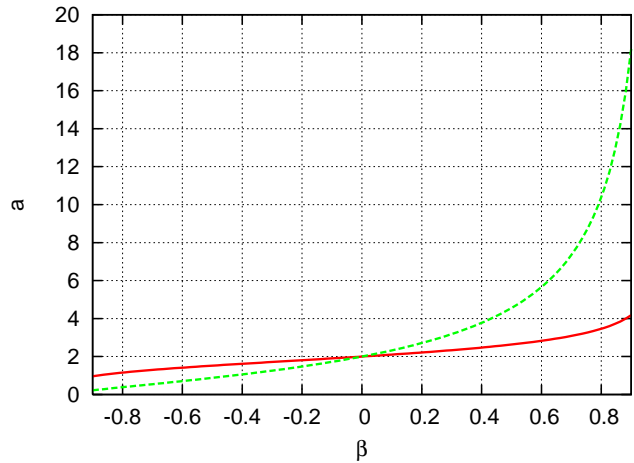,width=0.5\linewidth,angle=-0}}
\caption{{\protect\footnotesize {(Color online)The plot for the
solitary wave amplitude $a_{1,2}$ dependence on parameter $\beta$.
Red solid curve corresponds to the amplitude of the pulse in the
nonlinear PR channel, dashed green curve is for the NR channel.}}}
\label{Fig1}
\end{figure}

Some remarks are worthy to be made about the free parameter $\beta$.
The negative value of the parameter $\beta$ corresponds to the
solitary wave propagating in the direction of the axis $\zeta$. The
solitary wave characterized by positive value of the parameter
$\beta$ propagates in the opposite direction. The large amplitudes
(more powerful solitary waves) correspond to large positive values
of the parameter $\beta$. For the negative values of the parameter
$\beta$ the quasisolitons with smaller amplitudes have smaller
values of the parameter $\beta$ however the absolute value of the
velocity determined by the parameter $\beta$ is larger for the less
powerful solitary waves. We will refer to the Fig.\ref{Fig1} later
in discussion the influence of the linear losses at the
quasisoliton.

\section{Numerical simulation}

\noindent  It is common knowledge that the completely integrable
evolution nonlinear equations have a special solutions describing
elastic interaction of solitary waves \cite{Ablo}. These waves are
named solitons. It is our opinion that the system of equations
(\ref{eq2}) does not belong to the class of completely integrable
equations. Hence the solution of these equations does not represent
true soliton. We will denote them as quasisolitons. To investigate
interaction between the steady state solitary waves (\ref{eq13a})
and (\ref{eq13b}) the numerical simulation was pursued.

Numerical solution of the system of equations (\ref{eq2}) is
performed using the
 scheme of the finite differences. The conservation of the first integral of
motion (\ref{intmot}) is used to control the computational error. In
the consequent an integration step over the evolution variable
$\tau$ is set to $h\tau=0.0005$. The nonlinearity is set to $r=1$.
(The smaller nonlinearities correspond to the quasisolitons with the
same profiles but the smaller amplitudes (defined by the parameter
$\Theta$ which is linearly proportional to the nonlinearity
coefficient.)

\begin{figure} \centerline{
\epsfig{file=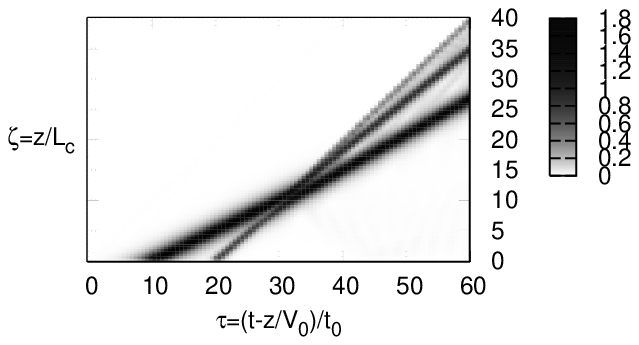,width=0.5\linewidth,angle=0}
\epsfig{file=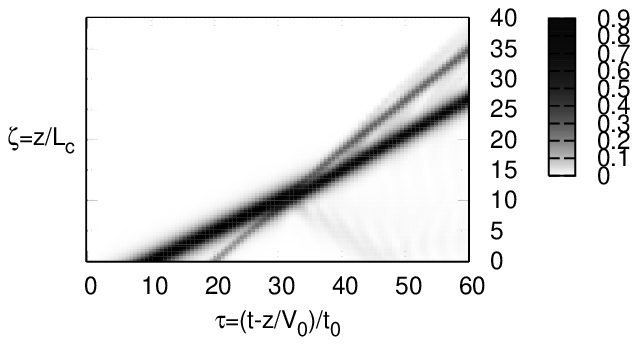,width=0.5\linewidth,angle=0} } \caption{
Crossing collision between two solitary waves with $\beta=-0.9$ and
$\beta=-0.5$. Left panel is for the solitary wave in the PR
waveguide, right panel is for the solitary wave in the NR waveguide.
} \label{m05m09}
\end{figure}

It was found that the results of the collision depend on the
relative velocity of the pulses. Fig.\ref{m05m09} illustrates
collision between quasisolitons characterized by $\beta=-0.9$ and
$\beta=-0.5$. The relative interaction velocity is $0.4$. The
quasisolitons drop some radiation after interaction and weakly
radiating wave could be noticed in the left panel of the picture as
a result of collision. Velocities and the amplitudes of the
resulting quasisolitons are slightly different from the parameters
of the quasisolitons before the collision due to loss of some part
of energy.

\begin{figure}
\centerline{
\epsfig{file=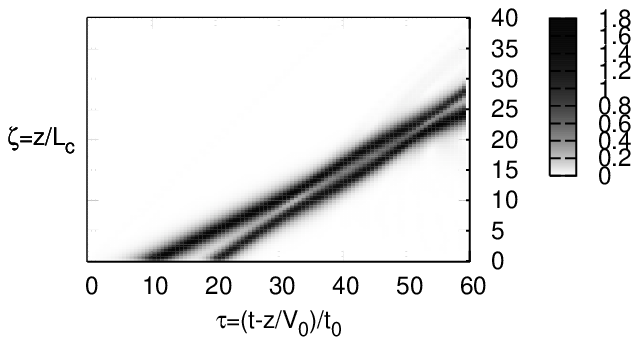,width=0.5\linewidth,angle=0}
\epsfig{file=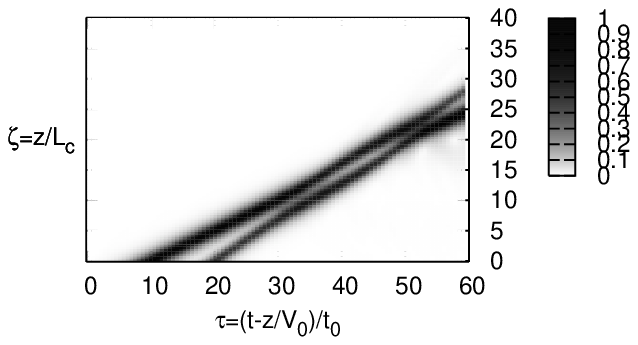,width=0.5\linewidth,angle=0} } \caption{
Repulsive collision between two solitary waves with $\beta=-0.7$ and
$\beta=-0.5$. Left panel is for the solitary wave in the PR
waveguide, right panel is for the solitary wave in the NR
waveguide.} \label{m05m07}
\end{figure}

Decrease the difference between velocities of colliding pulses
results in long-range interaction between pulses. The interaction
distance depends on the relation of the velocities. In the
Fig.\ref{m05m07} is shown interaction of two quasisolitons
characterized by $\beta=-0.7$ and $\beta=-0.5$. The relative
interaction velocity is $0.2$. The quasisolitons exchange energy at
the distance and form a transitory bound state.

It was mentioned before that the quasisolitions with the same
absolute values of velocity have different amplitudes. To
investigate the robustness of the quasisoliton depending on its
energy, at the Fig.\ref{m05m07} we present results of modeling the
collision between two steady-state solitary waves with the same
absolute values of the velocity $\beta=0.7$ and $\beta=-0.7$. A more
energetic quasisoliton with $\beta=0.7$ remains unchanged after
collision and the less energetic quasisoliton with $\beta=-0.7$
loose some radiation and changes its trajectory.
\begin{figure}
\centerline{ \epsfig{file=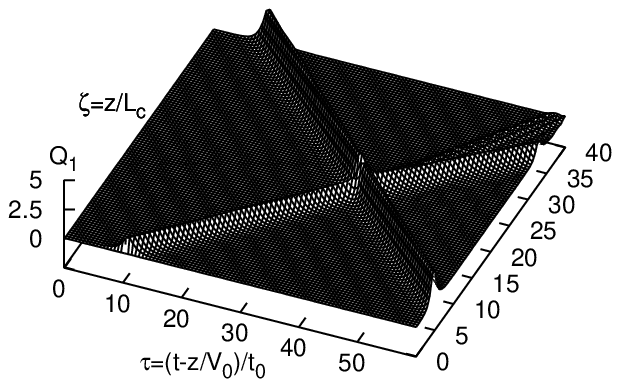,width=0.5\linewidth,angle=0}
\epsfig{file=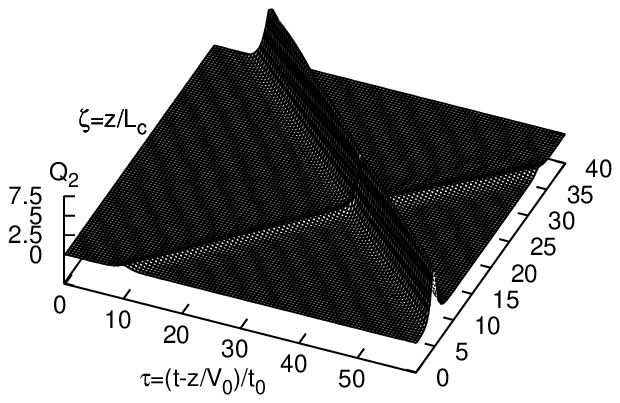,width=0.5\linewidth,angle=0} } \caption{
Collision between two solitary waves with  $\beta=0.7$ and
$\beta=-0.7$. Left panel is for the solitary wave in the PR
waveguide, right panel is for the solitary wave in the NR
waveguide.} \label{07m07}
\end{figure}

Thus, the collision of two steady state pulses with different
velocities has shown significant robustness. Small amplitude
radiation appearing after collision attests that the quasisoliton
eventually be disappeared.

For the current state of fabrication technology the losses in the
real NR materials are considerable. High value of losses renders
steady state pulse propagation impossible. Nevertheless, the role of
small losses would be considered. To check the influence of linear
losses in the NR channel on the solitary wave formation and
propagation, the model equations (\ref{eq2}) were modified by
including the additional term in the equation for a NR channel
\begin{eqnarray}
i \left(  \frac{\partial}{\partial
\zeta}+\frac{\partial}{\partial\tau}\right)Q_1+Q_2+r|Q_1|^2Q_1&=0  \label{eq14} \\
i \left(  \frac{\partial}{\partial
\zeta}-\frac{\partial}{\partial\tau}-\gamma\right)Q_2-Q_1&=0.
\end{eqnarray}

\begin{figure}
\centerline{ \epsfig{file=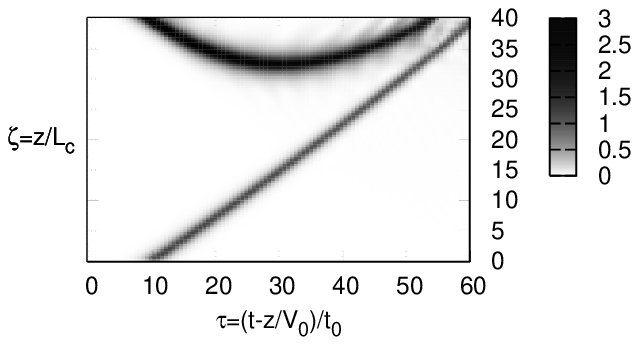,width=0.5\linewidth,angle=0}
\epsfig{file=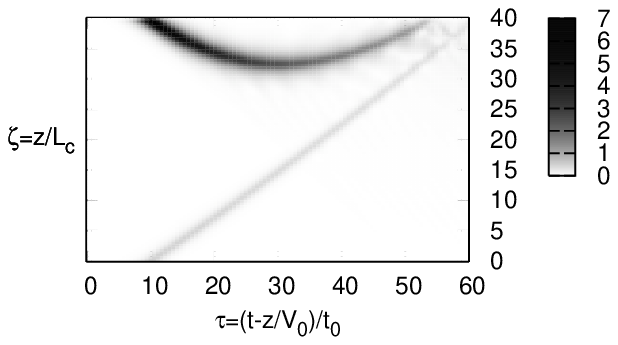,width=0.5\linewidth,angle=0} } \caption{
Two solitary waves with $\beta=0.7$ and $\beta=-0.7$ in a system
with the lossy NR channel  $\gamma=0.05$. Left panel is for the
solitary wave in the PR waveguide, right panel is for the solitary
wave in the NR waveguide.} \label{07m07dec}
\end{figure}

The Fig.\ref{07m07dec} illustrates the evolution of same pulses as
in the Fig.\ref{07m07} placed at the boundaries however the linear
losses in the NR channel are taken into consideration $\gamma=0.05$.
Comparing these two pictures one may conclude that even small losses
affect considerably the propagation properties of the solitary
waves. Propagating in the dissipative medium these pulses loose
energy and the pulse with positive $\beta$ slows down to sustain the
steady state regime. The pulse with negative $\beta$ slightly
accelerates to support the quasi-steady regime of propagation
(notice a Fig.\ref{Fig1}).

\section{Conclusion}

\noindent We considered the nonlinear solitary waves propagating in
the nonlinear opposite-directed coupler. One of its components is a
nonlinear waveguide made from the material with positive index of
refraction. Another channel is fabricated from linear dielectric
material with negative refraction index. For the system of coupled
waves the analytical solution for the electromagnetic waves in the
form of steady state solitary waves is found. It is interesting to
notice that the wave in the nonlinear channel affects the wave
propagation in the neighboring linear channel. This influence
results in coupling and a steady propagation of the solitary waves
in both waveguides, the wave in the nonlinear waveguide draws the
wave in the linear NR waveguide.

It was shown that the result of interaction of the quasisolitons
depends on their velocities. Provided that relative velocity of the
two colliding quasisolitons is large, the pulses collide almost like
the solitions. A weak radiation or a weak pulse emerges as a result
of collision. At small relative velocities of the quasisolitons the
interaction between two colliding solitary waves could results in a
strong energy exchange between colliding solitary waves and
formation of the temporarily coupled state of the interacting
quasisolitons. This phenomena is known for the case of nonlinear
directed coupling. In the case under consideration the formation of
the coupled state demands for an additional investigation. It could
be possible that a long-living coupled state of two solitary waves
is exists there.

The influence of the linear losses in the NR waveguide on the
existence of the solitary waves was studied. The wave loses its
energy as it propagates through the dissipative medium. In order to
compensate the losses the entering pulse in a form of the solitary
wave characterized by positive value of parameter $\beta$ (its
propagation direction is opposite to the coordinate axis $\zeta$)
permanently transforms into another solitary wave, less energetic,
having smaller value of the parameter $\beta$. It results in slowing
down the wave in the lossy medium. After that the wave stops in the
waveguide and changes the direction of its propagation. Oppositely
the entering pulse in a form of the solitary wave characterized by
negative parameter $\beta$ (its propagation direction is coincides
to the coordinate axis $\zeta$) tends to increase its velocity (i.e.
to decrease parameter $\beta$ which is negative) to sustain the
steady-state regime of propagation in presence of the losses.

\section*{Acknowledgments}

\noindent We are pleased to thank our colleagues Prof. S.O. Elyutin
and Prof. I.R. Gabitov for useful discussions. A.I. Maimistov
appreciates the support and hospitality of the Department of
Mathematics in the University of Arizona during his work under this
paper. The research of A.I. Maimistov and S.S. Ozhenko was partially
supported by RFBR(N 06-02-16406).

\end{document}